\documentclass[epsfig,12pt]{article}

\usepackage{epsfig}

\newcommand{\be}{\begin{equation}}
\newcommand{\ee}{\end{equation}}
\newcommand{\ba}{\begin{eqnarray}}
\newcommand{\ea}{\end{eqnarray}}
\newcommand{\bd}{\begin{displaymath}}
\newcommand{\ed}{\end{displaymath}}

\def\thalf{{\textstyle{\frac{1}{2}}}}

\def\root{\sqrt{-g}}
\def\Dr{\frac{d}{dr}}

\title{
{\bf Shear Transport Coefficients from Gauge/Gravity Correspondence}}
\author{{J. I. Kapusta and T. Springer} \vspace*{0.1in}\\
{\it School of Physics and Astronomy}\\
{\it University of Minnesota}\\
{\it Minneapolis, Minnesota 55455, USA}}

\date{August 11, 2008}

\begin{document}

\maketitle

\begin{abstract}
We study the shear mode in the gauge/gravity correspondence at finite temperature.  First, we confirm the general formula for the shear viscosity in an arbitrary background metric which includes a black hole in the fifth dimension.  We then derive a general formula for the shear mode relaxation time which appears in the theory of relativistic dissipative fluid dynamics; it agrees with known expressions in the limit of conformal fields.  These results may be useful in relativistic viscous fluid descriptions of high energy nuclear collisions at RHIC and LHC.
\end{abstract}

\vspace{0.5cm} {PACS numbers: 11.25.Tq, 12.38.Lg, 12.38.Mh, 47.75.+f}

\newpage

\section{Introduction}

The AdS/CFT correspondence \cite{Maldacena} or, more generally, the
relationship between gauge theories and higher dimensional gravity
theories, is a useful tool for studying strongly coupled plasmas, such
as the fluid created at RHIC (Relativistic Heavy Ion Collider) during
heavy ion collisions.  One can extract hydrodynamic transport
coefficients of the plasma by examining perturbations in the dual
gravity theory.  Usually this is done by either invoking the AdS/CFT
prescription to compute correlation functions of the stress-energy
tensor \cite{Policastro1, recipe, hydroI, hydroII, Herzog1} or by
applying appropriate boundary conditions and examining the resulting
dispersion relation \cite{Kovtun1, Mas}. 
In \cite{stretched}, the shear viscosity of currents defined on a
black hole's stretched horizon was computed for a general class of
supergravity backgrounds.  The value of the stretched horizon shear
viscosity agrees in all cases with the shear viscosity computed via
the methods previously listed \cite{Membrane}.  Though this agreement is not
understood at present, this membrane paradigm approach provides yet
another method for computing the transport coefficients of the
strongly coupled plasma (assuming that the agreement between the membrane
paradigm and AdS/CFT approaches is absolute).  We employ the membrane
paradigm approach in this work.

Hydrodynamics is an effective theory which describes the dynamics of a
thermal system on length and time scales much larger than any
microscopic scale.  Here we consider perturbations of the fluid in the
hydrodynamic regime; in this case, the energy and momentum of the
perturbations must be much less than the temperature of the plasma
$\omega , q \ll T$.  In a hydrodynamic theory, the spatial components
$T^{ij}$ of the stress-energy tensor are constructed from the
conserved quantities $T^{00} \equiv \epsilon$ (energy density) and
$T^{0i} \equiv \pi^i$ (momentum density).  These `constitutive
relations' arise by writing all possible terms consistent with the
symmetries present, taking into account equilibrium thermodynamics,
and by ignoring higher derivative terms since the hydrodynamic modes
are assumed to be slowly varying in space and time.

As an example, consider a translationally invariant, four dimensional
theory in flat space.  To first order in the perturbations and derivatives the constitutive relations are 
\be 
T^{ij} =
\delta^{ij}(P + v_s^2 \delta \epsilon) - \frac{\zeta}{w} \delta^{ij}
\partial_k \pi^k - \frac{\eta}{w} 
\left(\partial^i \pi^j + \partial^j\pi^i - \frac{2}{3} \delta^{ij} \partial_k \pi^k \right) \, .  
\ee
Here the Latin indices $i,j$ run over the 3 spatial coordinates.  The symbols $P$ and $w$ represent the equilibrium pressure and enthalpy density,
respectively, and $v_s$, $\eta$ and $\zeta$ represent the speed of sound,
shear and bulk viscosities.  Of course the energy-momentum tensor obeys the conservation equation, with the Greek indices $\mu,\nu$ running over all four coordinates, 
\be 
\partial_{\mu}T^{\mu \nu} = 0 \, .  
\ee 

These linearized hydrodynamic equations admit two normal modes,
corresponding to whether the momentum density fluctuations are
transverse or longitudinal to the fluid flow.  Transverse fluctuations
lead to the \emph{shear mode} which is the one considered in this
paper.  There is also the \emph{sound mode} (from longitudinal
momentum fluctuations), and the \emph{diffusive mode} (in the presence
of a conserved current); these modes are not considered in this work.
For a more complete introduction to relativistic hydrodynamics see \cite{Kovtun2}. 

Hydrodynamic fluctuations in the gauge theory plasma are dual to perturbations of a specified gravitational background in the dual theory.  Throughout this paper we use an effective 5-dimensional metric.  In principle, one can view this metric as being derived by dimensionally reducing some theory in a higher dimension.  Our results are easily generalizable to different numbers of dimensions in the same way as the results of \cite{stretched}.  Our convention
is to use $t,x,y,z$ to denote the usual four space-time coordinates; the coordinate $r$ denotes the extra dimension.

In the case of the shear mode, one only needs to consider metric perturbations $g_{\mu \nu} \rightarrow g_{\mu \nu} + h_{\mu \nu}$.  The components of the perturbation $h_{\mu \nu}$ can be decomposed into three irreducible sets; the field equations for each set decouple from the others.  The sets are found by classifying the perturbations under $O(2)$ rotations \cite{hydroI}; see also \cite{Mas}.  The set of gravitational perturbations dual to the hydrodynamic shear mode have $h_{y0}$, $h_{yz}$, and $h_{yr}$ all nonzero with all other components of $h_{\mu \nu}$ vanishing.  We assume a standard gauge choice $h_{\mu r} = 0$ which leaves us with only two nonvanishing components of $h_{\mu \nu}$.

The hydrodynamic transport coefficients are found by solving the linearized Einstein equations for the perturbations with appropriate boundary conditions.  The resulting dispersion relation for these gravitational perturbations can be compared with the expected dispersion relation from the boundary (gauge) hydrodynamics.  In first order hydrodynamics, this method has been used to calculate the speed of sound and the shear and bulk viscosities in a wide variety of gravity duals.  In \cite{stretched}, the authors also derive a general formula for the shear viscosity which depends only on components of the dual metric; this formula is applicable for a large class of gravitational backgrounds.

Recently, several groups have begun to extend these computations to second order hydrodynamics \cite{conformalrelax1, conformalrelax2, Heller1}.  Currently, there is no unique formalism for second order hydrodynamics, though a frequently used one is the ``M\"{u}ller-Israel-Stewart" formalism \cite{MIS}.  All such extensions of first order hydrodynamics attempt to repair problems that the first order theory has with causality, and necessarily introduce a set of new transport coefficients.  One such quantity is the ``shear relaxation time" $\tau$, and appears in the next to leading order in the shear dispersion relation, as explained in Sec. 5 below.  

In this paper we derive a general formula for the shear relaxation time analogous to the formula for the shear viscosity given in \cite{stretched}.  This formula is applicable to a wide variety of gravitational duals, and reproduces all known results in the literature of $\tau$.

This paper is organized as follows. In Sec. 2 we define the relevant gravitational background, the background Einstein equations, and the conservation equations that must be satisfied by the perturbed stress-energy tensor.  Our general relativistic conventions are those of \cite{Weinberg}.  In Sec. 3 we set up the gravitational shear perturbations, and determine the necessary linearized Einstein equations and boundary conditions that must be satisfied by the perturbations.  In Sec. 4 we proceed to solve these equations to lowest nontrivial order in $q$.  In doing so, we reproduce the formula for the shear viscosity which was derived by a different method in
\cite{stretched}.  In Sec. 5 we extend the calculation to the next order in $q$, thus deriving the corresponding general formula for the relaxation time $\tau$.  In Sec. 6 we consider applications of this result to certain phenomenological models of QCD, namely, the ``Hard Wall" and ``Soft Wall".  We make some conclusions and mention some prospects for further work in Sec. 7.   

\section{Metric Perturbations}

We consider a background metric of the generic form
\be
ds^2 = g_{00}(r) dt^2 + g_{xx}(r) |d{\bf x}|^2 + g_{rr}(r) dr^2
\label{backgroundg}
\ee
where the three functions $g_{00}$, $g_{xx}$ and $g_{rr}$ depend on the fifth dimensional variable $r$ only.  Our convention is such that $g_{00} = - 1$ and $g_{xx} = 1$ in flat 4-dimensional space.  The resulting Riemann-Christoffel curvature tensor and Ricci tensor are given in the appendix.  We assume that there is a horizon located at $r = r_0$.  As the horizon is approached we assume that the metric behaves as follows: $g_{rr} \rightarrow \gamma_r/(r-r_0)$ and $g_{00} \rightarrow - \gamma_0 (r-r_0)$ while $g_{xx}$ approaches the finite value $g_{xx}(r_0)$.  Thus $\root \rightarrow \sqrt{\gamma_0 \gamma_r g_{xx}^3(r_0)}$.  The Hawking temperature is
\be
T = \frac{1}{4\pi} \sqrt{\frac{\gamma_0}{\gamma_r}} \, .
\ee

The matter and radiation fields which generate this metric are not known.  Nevertheless, the energy-momentum tensor which generates it can be inferred from the Einstein equations.  They can be written either as
\be
G_{\mu\nu} = R_{\mu\nu} - \thalf R g_{\mu\nu} = -8\pi G_d T_{\mu\nu}
\ee
or as
\be
R_{\mu\nu} = -8\pi G_d \left( T_{\mu\nu} - \frac{1}{d-2} T^{\lambda}_{\;\;\lambda} g_{\mu\nu} \right)
\ee
where the number of space-time dimensions is $d$ and $G_d$ is the corresponding Newton constant.  In this paper we focus on $d = 5$.  With the notation used in the appendix this yields explicit expressions for the background energy-momentum tensor,
\ba
T_{00} = \frac{g_{00}}{16\pi G_5} \left( - F_0 + 3 F_x + F_r \right) 
\, , \nonumber \\
T_{xx} = T_{yy} = T_{zz} = \frac{g_{xx}}{16\pi G_5} \left( F_0 + F_x + F_r \right) \, , \nonumber \\
T_{rr} = \frac{g_{rr}}{16\pi G_5} \left( F_0 + 3 F_x - F_r \right) \, .
\label{backgroundT}
\ea
For a space with constant negative curvature $F_0 = F_x = F_r = 4/L^2$ and $R = 20/L^2$.  Then $T_{\mu\nu} = (3/4 \pi G_5 L^2) g_{\mu\nu}$.

For metric perturbations $g_{\mu\nu} \rightarrow g_{\mu\nu}(r) + h_{\mu\nu}(t,{\bf x},r)$ we allow for a perturbation in the energy-momentum tensor which is first order in $h_{\mu\nu}$ and its derivatives.
\be
T^{\mu\nu} = T^{\mu\nu}_{(0)} + T^{\mu\nu}_{(1)}.
\ee
Here $T^{\mu\nu}_{(0)}$ satisfies the Einstein equations with the background metric $g_{\mu\nu}$ given in Eq. (\ref{backgroundg}), see Eq. (\ref{backgroundT}), and $T^{\mu\nu}_{(1)}$ is its correction to first order in $h_{\mu\nu}$.  In addition it must satisfy the equation for conservation of energy and momentum.  Expanding the exact equation
\be
T^{\mu\nu}_{\;\;\;\; ;\mu} = 0
\ee
to first order one gets
\be
T^{\mu\nu}_{(1);\mu} + 
T^{\nu\lambda}_{(0)} \left( \delta \Gamma \right)^{\mu}_{\mu\lambda}
+ T^{\mu\lambda}_{(0)} \left( \delta \Gamma \right)^{\nu}_{\mu\lambda} = 0
\ee
where $\left( \delta \Gamma \right)^{\nu}_{\mu\lambda}$ is the change in the affine connection to first order in $h_{\mu\nu}$.  For a space of constant negative curvature, as in a conformal field theory with a gravity dual, the answer is easy to obtain.  It is just $T^{\mu\nu}_{(1)} = (3/4 \pi G_5 L^2)  h^{\mu\nu}$.  In the general case it is more involved.

\section{Shear Modes}

We consider a metric fluctuation of the following form:
\ba
h_{y0}(z,t,r) &=& g_{xx}(r) A(r) {\rm e}^{i(qz-\omega t)} \, , \\
h_{yz}(z,t,r) &=& g_{xx}(r) B(r) {\rm e}^{i(qz-\omega t)} \, .
\label{hyz}
\ea
Fluctuations in the metric are intimately connected with fluctuations in the energy-momentum tensor density.  For linear response it is natural to assume that $T^{\mu\nu}_{(1)}$ has the same harmonic variation in $z$ and $t$.  The conservation equations then take the following form:
\be
-i\omega T^{00}_{(1)} + iq T^{z0}_{(1)}
+ \frac{1}{\root g_{00}} \Dr \left( \root g_{00} T^{r0}_{(1)} \right) = 0 \, ,
\ee
\be
-i\omega T^{0x}_{(1)} + iq T^{zx}_{(1)}
+ \frac{1}{\root g_{xx}} \Dr \left( \root g_{xx} T^{rx}_{(1)} \right) = 0 \, ,
\ee
\ba
-i\omega T^{0y}_{(1)} + iq T^{zy}_{(1)}
+ \frac{1}{\root g_{xx}} \Dr \left( \root g_{xx} T^{ry}_{(1)} \right) 
\nonumber \\
-i\omega T^{00}_{(0)} g^{xx} h_{0y} 
+iq T^{xx}_{(0)} g^{xx} h_{zy} = 0 \, ,
\ea
\be
-i\omega T^{0z}_{(1)} + iq T^{zz}_{(1)}
+ \frac{1}{\root g_{xx}} \Dr \left( \root g_{xx} T^{rz}_{(1)} \right) = 0 \, ,
\ee
\ba
-i\omega T^{0r}_{(1)} + iq T^{zr}_{(1)}
+ \frac{1}{\sqrt{-g g_{rr}}} \Dr \left( \sqrt{-g g_{rr}} T^{rr}_{(1)} \right) 
-\thalf g^{rr} \frac{dg_{00}}{dr} T^{00}_{(1)}
\nonumber \\ 
-\thalf g^{rr} \frac{dg_{xx}}{dr} \left( T^{xx}_{(1)}+ T^{yy}_{(1)}+
T^{zz}_{(1)} \right) = 0 \, .
\ea
These equations are naturally satisfied by the choice
\be
T^{(1)}_{\mu\nu} = \frac{1}{16 \pi G_5} \left( F_0 + F_x + F_r \right)
h_{\mu\nu}
\ee
for arbitrary values of $\omega$ and $q$.  This is a particular solution of the inhomogeneous equations, that is, for given $h_{y0}$ and $h_{yz}$.  To it one may add solutions to the homogeneous equations, but this is not necessary to obtain the shear dispersion relation.

The above analysis leads to three nontrivial Einstein equations of motion,
\ba
R^{(1)}_{y0} &=& F_x h_{y0} \, ,\\
R^{(1)}_{yz} &=& F_x h_{yz} \, , \\
R^{(1)}_{yr} &=& 0 \, .
\ea
They can be expressed in terms of the unknown functions $A(r)$ and $B(r)$,
\be
\frac{1}{\root} \Dr \left[ \root g^{rr} g^{00} g_{xx}
\frac{dA}{dr}\right]
= q g^{00} \left( q A + \omega B\right) \, ,
\label{hyzA}
\ee
\be
\frac{1}{\root} \Dr \left[ \root g^{rr} \frac{dB}{dr}\right]
= \omega g^{00} \left( q A + \omega B\right) \, ,
\label{hyzB}
\ee
\be
\omega g^{00} \frac{dA}{dr} = q g^{xx} \frac{dB}{dr} \, .
\label{hyzC} 
\ee
Of course these cannot all be independent since there are only two functions $A$ and $B$.  Substitution of Eq. (\ref{hyzC}) into Eq. (\ref{hyzB}) reproduces Eq. (\ref{hyzA}), for example.

A boundary condition is needed near the horizon.  From the structure of the equations near the horizon we observe that $A \sim (r-r_0)B$ as $r \rightarrow r_0$.  In this limit Eq. (\ref{hyzB}) implies that
\be
\frac{\partial^2}{\partial t^2} h_{yz} = \frac{\gamma_0}{\gamma_r} (r-r_0) 
\frac{\partial}{\partial r} \left[ (r-r_0) \frac{\partial}{\partial r}
h_{yz} \right]
\ee
which has two independent solutions corresponding to waves incoming to the horizon and outgoing from the horizon,
\be
h_{yz} = f_{\rm in}\left(t+\sqrt{\frac{\gamma_r}{\gamma_0}} \ln(r-r_0) \right)
+ f_{\rm out}\left(t-\sqrt{\frac{\gamma_r}{\gamma_0}} \ln(r-r_0) \right) \, .
\ee
The boundary condition is to select only the incoming waves.  This results in the following requirement:
\be
\frac{\partial h_{yz}}{\partial t} = \sqrt{\frac{\gamma_0}{\gamma_r}}
(r-r_0) \frac{\partial h_{yz}}{\partial r} \,\,\,\, {\rm as} \,\,\,\,
r \rightarrow r_0 \, .
\ee
For the exponential time dependence assumed in Eq. (\ref{hyz}) this is
\be
-i\omega h_{yz} = \sqrt{\frac{\gamma_0}{\gamma_r}}
(r-r_0) \frac{\partial h_{yz}}{\partial r} \,\,\,\, {\rm as} \,\,\,\,
r \rightarrow r_0 \, .
\label{BChyz}
\ee
The idea proposed in \cite{stretched} is to apply this condition at a distance $r_h$ such that $r_h -r_0 \ll r_0$, which is referred to as a stretched horizon.  This is then inserted on the left side of Eq. (\ref{hyzB}) to obtain
\be
(r-r_0) \frac{dB}{dr} = -i\sqrt{\frac{\gamma_r}{\gamma_0}}
\left( qA + \omega B \right) \,\,\,\, {\rm as} \,\,\,\, r \rightarrow r_h
\, .
\label{stretchedBC}
\ee
The reason for introducing the stretched horizon has to do with the dispersion relation $\omega = -iD q^2 +\cdot \cdot \cdot$.  Although $|A| \ll |B|$ as $r \rightarrow r_0$, consideration of very small but nonvanishing wave-number $q$ will result in $q|A| \sim |\omega B|$ and eventually $q|A| \gg |\omega B|$ for a fixed value of $r$ as $q \rightarrow 0$.  The latter can be satisfied self-consistently if $r_h-r_0$ is not too small.  A more precise condition will be given below.
 
\section{Shear Viscosity}

The shear modes are transverse and strongly over-damped.  At low frequency and wave-number they have the dispersion relation $\omega = -iD q^2$ where $D$ is a diffusion constant.  In terms of the shear viscosity $\eta$ and enthalpy density $w = Ts = P+\rho$ it is $D = \eta/w$.  Upon inspection, and with the expectation of a diffusive mode, we expand the functions in powers of $q$ as follows:
\ba
A(r) &=& \sum_{n=0}^{\infty} q^{2n} A^{(2n)}(r) \, , \\ \nonumber
B(r) &=& \sum_{n=0}^{\infty} q^{2n+1} B^{(2n+1)}(r) \, .
\ea
From Eqs. (\ref{hyzA}) and (\ref{hyzB}) we find the lowest order solutions
\ba
A^{(0)}(r) &=& a_0 + a_1 \int_{\infty}^r dr' \frac{g_{rr}(r') g_{00}(r')}
{\sqrt{-g(r')} g_{xx}(r')} \, , \\
B^{(1)}(r) &=& b_0 + b_1 \int_{\infty}^r dr' \frac{g_{rr}(r')}
{\sqrt{-g(r')}} \,.
\ea
Substitution into Eq. (\ref{hyzC}), together with $\omega = - i D q^2$, gives the relationship $D = i b_1/a_1$.  The Dirichlet boundary condition on $A$ and $B$ at infinity gives $a_0 = b_0 = 0$.

It is now a straightforward matter to substitute the lowest order solution into the boundary condition on the stretched horizon to obtain the shear diffusion constant,
\be
D = \frac{\sqrt{-g(r_0)}}{\sqrt{-g_{00}(r_0) g_{rr}(r_0)}}
\int_{r_0}^{\infty} dr \frac{-g_{00}(r) g_{rr}(r)}{\sqrt{-g(r)} g_{xx}(r)} \, .
\label{diffusion}
\ee
Here the limit $r_h \rightarrow r_0$ has been taken without any harm.  This is identical to the result obtained in \cite{stretched} in which Maxwell's equations were solved (in the specified metric) rather than Einstein's equations.

A frequently used metric is one that is dual to ${\cal N} = 4$ SU($N$) supersymmetric Yang Mills theory at finite temperature and in the limit $N \rightarrow \infty$ and $g^2 N \rightarrow \infty$,
\be
ds^2 = \frac{r^2}{L^2} \left[ -f(r)dt^2 + |d{\bf x}|^2 \right]
 + \frac{L^2}{r^2 f(r)} dr^2 \, .
\label{SUSY}
\ee   
Here $f(r) = 1-(r_0/r)^4$.  Using this metric yields
\ba
D &=& \frac{1}{4\pi T} \, , \nonumber \\
\frac{\eta}{s} &=& \frac{1}{4\pi} \, ,
\label{ConformalEta}
\ea
which are well-known results.
 
The self-consistency of the assumption that $|\omega B|$ could be neglected compared to $q|A|$ near the stretched horizon can now be checked.  It requires the following:
\be
r_0 \exp\left( -\frac{4\pi T}{D q^2}\right) \ll r_h - r_0 \ll r_0 \, .
\ee
Obviously there is a finite range of $q$ for which this condition can be met.  Nevertheless the derived expression for the diffusion constant and the ratio of shear viscosity to entropy density remains valid.

\section{Shear Relaxation Time}

The lowest order dispersion relation for shear modes is $\omega = -iDq^2$.  The partial differential equation which describes such a mode is the diffusion equation
\be
\frac{\partial}{\partial t} \psi - D \frac{\partial^2}{\partial z^2} \psi = 0
\ee
where $\psi$ represents a component of the shear-stress tensor.  The frequency is obviously an expansion in even powers of the wavenumber.  The next correction to the shear dispersion relation given above is fourth order in $q$.  If we want to write a partial differential equation to describe the shear mode we should add a term proportional to the fourth spatial derivative of the field.  For consistency, there ought to be a term proportional to the second temporal derivative of the field as well.  Therefore we are led to consider the partial differential equation
\be
\tau_1 \frac{\partial^2}{\partial t^2} \psi + \frac{\partial}{\partial t} \psi - D \frac{\partial^2}{\partial z^2} \psi + 
\tau_2 D^2 \frac{\partial^4}{\partial z^4} \psi = 0
\label{4PDE}
\ee
where two time constants $\tau_1$ and $\tau_2$ have been introduced.  This leads to a shear dispersion relation
\be
\omega = -iDq^2 \left( 1 + \tau D q^2 + \cdot \cdot \cdot \right)
\ee
where only the combination $\tau \equiv \tau_1 + \tau_2$ appears.  We refer to $\tau$ as a ``shear relaxation time" because it is associated with the shear mode and it has the dimensions of time.  The term is somewhat of a misnomer because it does not imply that fields falloff as $\exp(-t/\tau)$; see Sec. 6.  In \cite{conformalrelax1} the quantity which we denote as $\tau$ was denoted $\tau_\Pi$ and was believed to coincide with the ``relaxation time" in the M\"{u}ller-Israel-Stewart formulation of second order fluid dynamics \cite{MIS}.  However, in \cite{conformalrelax2} it was argued that the quantity we denote as $\tau$ contains not only the M\"{u}ller-Israel-Stewart $\tau_\Pi$, but also contributions from (currently unformulated) third order fluid dynamics.  Actually this is quite apparent from Eq. (\ref{4PDE}).  In this way the authors of \cite{conformalrelax2} explained the apparent discrepancy that arises when inferring $\tau_\Pi$ from the shear mode versus the sound mode. 
To clarify, we emphasize that our $\tau$ is the same quantity that is computed in \cite{conformalrelax1}, but is not the same as the $\tau_\Pi$ of M\"{u}ller, Israel and Stewart.  However, it is necessary input for the higher order relativistic dissipative fluid dynamic description of the flow of matter and radiation.  We shall now derive a formula for $\tau$.

The lowest order solution, which was all that was needed to determine the shear viscosity, was
\ba
A^{(0)}(r) &=& a_1 \int^{\infty}_r dr' \frac{-g_{00}(r') g_{rr}(r')}
{\sqrt{-g(r')} g_{xx}(r')} \, , \nonumber \\
B^{(1)}(r) &=& -iD a_1 \int^{\infty}_r dr' \frac{g_{rr}(r')} 
{\sqrt{-g(r')}} \,.
\ea
Substituting these into Eqs. (\ref{hyzA})-(\ref{hyzB}) leads to the next higher order terms,
\bd
A^{(2)}(r) = a_2 + \int_{\infty}^r dr' \frac{g_{rr}(r') g_{00}(r')}
{\sqrt{-g(r')} g_{xx}(r')} \left\{ a_3 + \int_{\infty}^{r'} dr''
\sqrt{-g(r'')} g^{00}(r'') A^{(0)}(r'') \right\} \, ,
\ed
\be
B^{(3)}(r) = b_2 + \int_{\infty}^r dr' \frac{g_{rr}(r')}
{\sqrt{-g(r')}} \left\{ b_3 - iD \int_{\infty}^{r'} dr''
\sqrt{-g(r'')} g^{00}(r'') A^{(0)}(r'') \right\} \, .
\ee
Dirichlet boundary conditions at infinity require $a_2 = b_2 = 0$.  Finally, substitution of these solutions into Eq. (\ref{hyzC}) leads to
\be
D^2 \tau = - D \frac{a_3}{a_1} + i \frac{b_3}{a_1} \, .
\ee

To complete the calculation the boundary condition on the stretched horizon (\ref{stretchedBC}) must be satisfied.  To this order it is
\be
(r-r_0) \frac{dB^{(3)}}{dr} = - \frac{i}{4\pi T}
\left( A^{(2)} - i D B^{(1)} \right) \,\,\,\, {\rm as} \,\,\,\,
r \rightarrow r_h \, .
\ee
This precisely determines the shear relaxation time
\be
\tau = \frac{\sqrt{-g(r_0)}}{\sqrt{-g_{00}(r_0) g_{rr}(r_0)}}
\int_{r_0}^\infty dr \frac{g_{rr}(r)}{\sqrt{-g(r)}}
\left[ 1- \left( \frac{D(r)}{D(r_0)} \right)^2 \right]
\label{relaxation}
\ee
where
\be
D(r) \equiv \frac{\sqrt{-g(r)}}{\sqrt{-g_{00}(r)g_{rr}(r)}}
\int_r^\infty dr^\prime 
\frac{-g_{00}(r^\prime)g_{rr}(r^\prime)}
{\sqrt{-g(r^\prime)}g_{xx}(r^\prime)} \, .
\ee
In deriving Eq. (\ref{relaxation}) an integration by parts was performed to reduce the number of nested integrals.  Note that $D(r_0)$ is just the diffusion constant $D$.  The limit $r_h \rightarrow r_0$ has been taken since the integrals converge.

The formula given in Eq. (\ref{relaxation}) is the main result of this paper.  For the special metric (\ref{SUSY}) it gives
\be
\tau = \frac{1 - \ln 2}{2\pi T} 
\label{ConformalTau}
\ee
which was already derived in \cite{conformalrelax1,conformalrelax2}.
In the literature this metric is referred to as AdS$_5$(D3) for 5-dimensional anti--deSitter D3-brane.  The shear relaxation time has been computed for other conformal theories as well, including AdS$_4$(M2) and AdS$_7$(M5).  The formula (\ref{relaxation}) reproduces all of these known results.

\section{Hard and Soft Wall Models of QCD}

There are several phenomenological models for QCD that incorporate confinement.  The simplest of these is the hard wall model.  A better one is the soft wall model as it represents the linear radial Regge trajectories for the families of $\rho$ and $a_1$ mesons quite well.  Here we compute the ratio of shear viscosity to entropy density and the shear relaxation time for these models since they may be useful in applications to high energy nuclear collisions at RHIC and LHC.

\subsection{Hard wall model}

The hard wall model \cite{hardwall} uses the metric of Eq. (\ref{SUSY}) but with $r$-space cutoff at some value $r_{\rm min}$.  This leads to radial excitations of the vector and axial-vector meson spectra with the mass being linear in the radial quantum number $n$.  Introduction of the new scale $r_{\rm min}$ implies that the transport coefficients do not scale dimensionally with temperature $T$ alone.

There is a significant difference depending on whether $r_{\rm min} < r_0$ or $r_{\rm min} > r_0$.  In the former case the introduction of a hard cutoff does not influence the calculations of the previous sections since it lies inside the black hole.  In the latter case it would seem to be indeterminate because the horizon lies beyond the wall and the boundary condition near the horizon cannot be implemented.  On the other hand, we can stretch the boundary out from $r=r_0$ to $r=r_{\rm min}$ if the distance $r_{\rm min} - r_0$ is so small that we can use the approximations $g_{00}(r) \approx -\gamma_0 (r-r_0)$ and $g_{rr}(r) \approx \gamma_r/(r-r_0)$.  Then the boundary condition on the stretched horizon which allows only for incoming, and not outgoing, waves can be applied.  The answers are the same as Eqs. (\ref{diffusion}) and (\ref{relaxation}) with $r_0$ replaced by $r_{\rm min}$.  One finds that
\be
\frac{\eta}{s} = \frac{1}{4\pi} \;\;\;\;\;{\rm for} \; T \ge T_c
\ee
and
\be 
\frac{\eta}{s} = \frac{1}{4\pi}\frac{T}{T_c} \;\;\;{\rm for} \; T \le T_c
\label{highTeta}
\ee
where the Hawking temperature is $T = r_0/\pi L^2$ and we have defined a critical hardwall temperature $T_c = r_{\rm min}/\pi L^2$.  If we are willing to tolerate a 5\% deviation of the metric from the limiting form given above then the expression (\ref{highTeta}) is only a good approximation for $0.9 \le T/T_c \le 1$. For lower temperatures it is difficult to cleanly separate the incoming from the outgoing waves.  The hard wall model is too crude to access lower temperatures.  Nevertheless it does give an indication of how $\eta/s$ might deviate from its conformal value when a confinement scale is introduced.  Note that it does fall below the conjectured lower bound for $\eta/s$ of $1/4\pi$ when $T < T_c$.

The hard wall result for the shear relaxation time is
\be
\tau = \frac{1 - \ln 2}{2\pi T} \;\;\;\;\; {\rm for} \; T \ge T_c
\ee
and  
\ba
\tau &=& \frac{1}{4\pi T} \left(\frac{T_c}{T}\right)^3
\Bigg\{ 2 - \left[ 1 + \left(\frac{T_c}{T}\right)^2 \right]
\ln\left[ 1 + \left(\frac{T}{T_c}\right)^2 \right] \nonumber \\ 
&-& \left[ 1 - \left(\frac{T_c}{T}\right)^2 \right]
\ln\left[ 1 - \left(\frac{T}{T_c}\right)^2 \right] \Bigg\}
 \;\;\;\;\; {\rm for} \; T \le T_c \, .
\ea
We expect this result to be reasonably accurate only when $0.9 \le T/T_c \le 1$ for the reasons mentioned above. In that case the formula can be expanded just below $T_c$ as follows:
\be
\tau \approx \frac{1}{2\pi T} \Bigg\{ 1 - \ln 2
- \frac{T_c-T}{T_c} \left[ \ln\left(\frac{8T_c}{T_c-T}\right) - 4 \right] \Bigg\} \, .
\ee
Note that the effect of the confinement scale is to reduce the shear relaxation time compared to the conformal limit.  Numerically, the value of $T_c$ can be estimated as follows.  The mass of the $\rho$-meson in the hard wall model is found by solving a particular wave equation whose solutions are Bessel functions.  A boundary condition requires that $J_0(m_{\rho}L^2/r_{\rm min}) = 0$ from which one deduces that $T_c \approx m_{\rho}/7.556 \approx 102$ MeV.  This is just a characteristic temperature, not the critical temperature of a phase transition, since one may define it somewhat differently \cite{softwallD}.

\subsection{Soft wall model}

The soft wall model \cite{softwall} was developed to improve upon the hard wall model.  In particular, it leads to linear Regge trajectories wherein the radial excitations of the vector and axial-vector meson spectra have mass-squared being linear in the radial quantum number $n$, in substantial agreement with data.  
It can be obtained by adding a dilaton field to the usual AdS metric. 
The piece of the action that is relevant for the meson mass spectra is 
\be 
S_{\rm meson}= -\frac{1}{4 g_5} \int d^5 x \sqrt{-g} e^{-\phi} 
(F_L^2 + F_R^2)
\label{softwallaction2}
\ee 
where $\phi$ is the dilaton field and $F_L^{\mu \nu}$ and $F_R^{\mu
\nu}$ are the field strength tensors for the left and right handed
gauge fields.  In our coordinates, the dilaton profile which leads to
the Regge behavior is $\phi(r) = c L^4/r^2$, where $c$ is a constant
which can be determined by fitting the meson spectrum.  If we compute the quantities $\eta/s$ and $\tau$, using the soft wall model of \cite{softwall}, the results are the same as Eqs. (\ref{ConformalEta}) and (\ref{ConformalTau}) because the metric is exactly $AdS_5$.  There is an alternative formulation of the soft wall.  Instead of adding a nontrivial dilaton, one keeps the dilaton constant while deforming the metric away from $AdS_5$.  This version of the soft wall model has been studied in \cite{softwallD2, Andreev1, Andreev2, Andreev3}.  The deformed metric is
\be 
d \hat{s}^2 = e^{-h(r)}\left[\frac{r^2}{L^2} \left( -f(r)dt^2 + |d{\bf x}|^2 \right) + \frac{L^2}{r^2 f(r)} dr^2 \right] \, .
\label{softwallmetric}
\ee
In order to preserve the essential feature of the soft wall model, namely, the
linear Regge trajectories, one must choose
\be
h(r) = 2\phi(r)
\ee 
so that the action (\ref {softwallaction2}) is the same with regard to the mesons.  We stress that this is a different implementation of the soft wall model than that of \cite{softwall}; it is not simply a transformation from the string frame to the Einstein frame, because the dilaton profile is modified as well.  Both pictures lead to linear Regge trajectories, but other physical quantities may differ, as we now demonstrate by computing $\eta/s$ and $\tau$.

Using this metric in Eq. (\ref{diffusion}) yields
\be
\frac{\eta}{s} = \frac{1}{2\pi x_0}\left[ 1+\frac{1}{x_0}
\left(e^{-x_0} -1 \right) \right] 
\ee
where, following \cite{softwallD} (see also \cite{softwallD2,Andreev1}),
we define a characteristic temperature via $T_c^2 \equiv c/4$ and where 
\be
 x_0 \equiv 12 \left( \frac{T_c}{\pi T} \right)^2 \, . 
\ee
In the high temperature limit
\be
\frac{\eta}{s} = \frac{1}{4\pi} \left[ 1 - 4 
\left( \frac{T_c}{\pi T} \right)^2 + 12 
\left( \frac{T_c}{\pi T} \right)^4 + \cdot\cdot\cdot \right]
\ee
while in the low temperature limit
\be
\frac{\eta}{s} = \frac{\pi}{24} \left(\frac{T}{T_c} \right)^2 
\left[ 1 - \frac{\pi^2}{12} 
\left( \frac{T}{T_c} \right)^2 + \cdot\cdot\cdot \right] \, .
\ee
Similar to the hard wall model, the conformal limit of $1/4\pi$ is approached from below.  See Fig. 1.  Unlike the hard wall model, the dependence of $\eta/s$ is smooth and well-defined for all temperatures.

One may be surprised that the shear viscosity is not greater than or equal to
$1/4\pi$ given the proof of \cite{Liu}.  The reason is that \cite{Liu} assumes a background metric with the property that $F_0(r) = F_x(r)$.  Instead, in this version of the soft wall model, one finds by explicit computation that
\be
F_0(r) - F_x(r) = -3 h^{\prime}(r)e^{h(r)}\frac{r_0^4}{r^3 L^2}
\ee  
which is nonzero as long as $h(r)$ is not constant.  Thus, we would
only expect the general proof of \cite{Liu} to apply in the case where
$h(r)$ is constant where the metric is pure $AdS_5$.
The fact that $F_0(r) \neq F_x(r)$ which implies $\eta/s \neq
1/4\pi$ may be a sign that a metric such as (\ref{softwallmetric}) cannot be generated by conventional supergravity matter fields.  However, as is customary, we view this metric as a phenomenologically motivated effective 5D theory, and thus will not concern ourselves with its origin in this paper.

The relaxation time can also be computed in the soft wall model.  Applying the formula (\ref{relaxation}) we find the relaxation time to be
\be
\tau = \frac{{\rm e}^{-x_0}}{2 \pi T}
\int_0^{x_0} dx \, \frac{x {\rm e}^x}{x_0^2-x^2}
  \left[1- \left(\frac{x_0}{x}\right)^3
  	\left(\frac{x + {\rm e}^{-x} -1}{x_0 + e^{-x_0} -1}\right)^2 
  \right] \, .
\ee
Although this integral may be expressed in terms of special functions, the result is not very enlightening.  The relaxation time is plotted in Fig. 2.  As for the hard wall model, the conformal limit is approached from below.  It is noteworthy that the shear relaxation time is negative for $T/T_c < 0.686$.  
Whether this is unphysical or not is discussed in the next subsection.
  
A high temperature expansion is easily derived.
\ba
\tau &=& \frac{1}{2\pi T}
\Bigg[ \left( 1 - \ln 2 \right) - 
2(7 - 8 \ln 2)\left(\frac{T_c}{\pi T} \right)^2
\nonumber \\
&+& \frac{4(91 - 120 \ln 2)}{3}\left(\frac{T_c}{\pi T} \right)^4 -
\frac{32(397 - 558 \ln 2)}{15} \left(\frac{T_c}{\pi T}\right)^6 + \cdot \cdot \cdot 
\Bigg]
\ea
This expansion is accurate to better than 10\% when $T/T_c > 0.9$.

Numerically, the value of $T_c$ can be estimated as follows.  The mass of the $\rho$-meson in the soft wall model is found by solving a particular wave equation whose solutions involve Laguerre polynomials.  A boundary condition leads to the relation $m_{\rho} = 2\sqrt{c}$.  The Hawking-Page analysis of the phase transition in \cite{softwallD} gives $T_c = m_{\rho}/4 \approx 192$ MeV.    
A similar analysis in the hard wall model yields a transition temperature of approximately $m_{\rho}/6.354$ \cite{softwallD}.

\subsection{Sign of the shear relaxation time}

Is a negative shear relaxation time unphysical?  First consider real, positive $q$ and complex $\omega$.  A shear plane wave propagating in the $z$-direction has the form 
\be
\psi = \psi_0 \, {\rm e}^{i(q z-\omega t)} = 
\psi_0 \, {\rm e}^{i(q z-\omega_R t)} {\rm e}^{\omega_I t}
\ee
where the frequency has been decomposed into its real and imaginary parts as $\omega = \omega_R + i \omega_I$.  The real part vanishes and the imaginary part is
\be
\omega_I = -Dq^2 \left( 1 + \tau D q^2 + \cdot \cdot \cdot \right)
\ee
which is definitely negative if $\tau > 0$. This is the usual situation where the waves falls off exponentially in time.  If $\tau < 0$ then the first two terms of the expansion cannot be trusted for $q^2 > 1/|\tau D|$.  But the first two terms of the expansion cannot be trusted for larger values of $q$ even if $\tau$ was positive. 

Now consider real, positive $\omega$ and complex $q$.  The shear dispersion relation is
\be
q^2 = \frac{\omega}{D} \left( i + \tau \omega + \cdot \cdot \cdot \right) \, .
\ee
The wavenumber can be decomposed into its real and imaginary parts as $q=q_R+iq_I$.  The dispersion relation to the required order is determined by the solution to the following equation:
\be
q_R^2 - q_I^2 - \frac{\tau \omega^2}{D} + i \left( 2 q_R q_I - \frac{\omega}{D}
\right) = 0 \, .
\ee
Obviously $q_R q_I = \omega/2D > 0$.  The shear wave behaves as
\be
\psi = \psi_0 \, {\rm e}^{i(q_R z-\omega t)} {\rm e}^{-q_I z} \, .
\ee
For a wave moving in the positive $z$-direction, $q_R > 0$ and $q_I > 0$, and the wave is damped in space.  For a wave moving in the negative $z$-direction, $q_R < 0$ and $q_I < 0$, and the wave is again damped in space.  Hence, these shear waves are damped in space for any finite value of the relaxation $-\infty < \tau < \infty$ no matter what its sign.

We are not aware of anything unphysical about a negative shear relaxation time according to the definition of it given in this paper.  
See also the analysis and discussion of this point in \cite{absorption}.  The only unpleasantness associated with a negative relaxation time may be numerical instabilities that could arise in relativistic viscous fluid dynamics due to finite numerical resolution. 

\section{Conclusions}

In this paper we have used the gauge/gravity correspondence at finite temperature to study the properties of the shear mode relevant to a four dimensional world.  We confirmed a general formula for the shear diffusion constant - equivalently the dimensionless ratio of shear viscosity to entropy density.  This quantity may be computed if the five dimensional metric is given.  It reproduces all known results in the literature.  The method we use allows us to determine the shear dispersion relation to arbitrary high order in the frequency and wavenumber.  In particular, the main result of this paper was to obtain a general formula for the coefficient of the next order term in the shear dispersion relation.  It has the dimension of time, and it is often referred to as a shear relaxation time, although that is somewhat of a misnomer as we explained earlier.  It too reproduces all known results in the literature.  We applied these formulas to models of confinement in QCD, namely, the hard and soft wall models.  These results may be useful in relativistic viscous fluid descriptions of high energy nuclear collisions at RHIC and LHC.

We are currently pursuing several avenues of research.  We would like to perform the same type of analysis for the sound mode.  However, that is a much more difficult undertaking, and seems to require more information than is necessary for the shear mode.  A precise criterion for determining which theories these results apply to has not yet been given in the literature.  We are investigating the coupling of arbitrary fields to the gravitational field in an AdS space to address both of these problems.

\section*{Acknowledgements}

We are grateful to the organizers and to the Institute for Nuclear Theory at the University of Washington for providing a stimulating atmosphere during the workshop ``String Theory Methods in the Real World" in May 2008.  
We would also like to thank Irene Amado, Oleg Andreev, Carlos Hoyos, Keijo Kajantie, Pavel Kovtun, Karl Landsteiner, Makoto Natsuume, and Andrei Starinets for helpful comments and feedback on this paper.  
This work was supported by the US Department of Energy (DOE) under Grant No.
DE-FG02-87ER40328.

\section*{Appendix}

The Riemann-Christoffel tensor $R_{\alpha \beta \mu \nu}$ for the background metric has the following independent components.  All the rest are either related to these by the algebraic properties of $R_{\alpha \beta \mu \nu}$ or are zero.
\ba
R_{r0r0} &=& \frac{1}{4} \frac{dg_{00}}{dr} \Dr \ln \left[ g^{rr} g^{00}
\left( \frac{dg_{00}}{dr} \right)^2 \right] \nonumber \\
R_{rxrx} = R_{ryry} = R_{rzrz} &=& \frac{1}{4} \frac{dg_{xx}}{dr} 
\Dr \ln \left[ g^{rr} g^{xx}
\left( \frac{dg_{xx}}{dr} \right)^2 \right] \nonumber \\
R_{0x0x} = R_{0y0y} = R_{0z0z} &=& \frac{1}{4} g^{rr} \frac{dg_{00}}{dr}
\frac{dg_{xx}}{dr} \nonumber \\
R_{xyxy} = R_{xzxz} = R_{yzyz} &=& \frac{1}{4} g^{rr} 
\left( \frac{dg_{xx}}{dr} \right)^2
\ea

The diagonal elements of the Ricci tensor $R_{\mu\nu} = g^{\alpha\beta}
R_{\alpha\mu\beta\nu}$ are nonzero while the off-diagonal ones are zero.
\ba
R_{00} &=& \frac{1}{2} g^{rr} \frac{dg_{00}}{dr} \Dr \ln \left[ \root
g^{rr} g^{00} \frac{dg_{00}}{dr} \right] \equiv g_{00} F_0 \nonumber \\
R_{xx} = R_{yy} = R_{zz} &=& \frac{1}{2} g^{rr} \frac{dg_{xx}}{dr} 
\Dr \ln \left[ \root g^{rr} g^{xx} \frac{dg_{xx}}{dr} \right] \equiv g_{xx} F_x \nonumber \\
R_{rr} &=& \frac{3}{4} g^{xx} \frac{dg_{xx}}{dr} \Dr \ln \left[ g^{rr} g^{xx}
\left( \frac{dg_{xx}}{dr} \right)^2 \right] \nonumber \\
&+& \frac{1}{4} g^{00} \frac{dg_{00}}{dr} \Dr \ln \left[ g^{rr} g^{00}
\left( \frac{dg_{00}}{dr} \right)^2 \right] \equiv g_{rr} F_r 
\ea

The curvature $R = R^{\lambda}_{\;\;\lambda}$ is given as follows:
\be
R = F_0 + 3F_x + F_r \, .
\ee

Lastly we come to the invariant
\be
R_{\mu \nu \lambda \sigma} R^{\mu \nu \lambda \sigma} = g^{\alpha \mu} g^{\beta
\nu} g^{\lambda \gamma} g^{\sigma \delta} R_{\alpha \beta \gamma \delta} R_{\mu \nu
\lambda \sigma} \, .
\ee
For the background metric this reduces to
\be
R_{\mu \nu \lambda \sigma} R^{\mu \nu \lambda \sigma} = 2 \sum_{\mu=0}^4 \sum_{\nu=0}^4 \left(g^{\mu\mu}g^{\nu\nu}
R_{\mu\nu\mu\nu}\right)^2
\ee
which can be written as
\ba
R_{\mu \nu \lambda \sigma}R^{\mu \nu \lambda \sigma} &=& 
\frac{1}{4} (g^{rr})^2\Bigg\{ 3 \left( g^{00} 
\frac{dg_{00}}{dr} g^{xx} \frac{dg_{xx}}{dr}\right)^2 + 3
\left(g^{xx}\frac{dg_{xx}}{dr}\right)^4  \nonumber \\
&+& 3 \left( g^{xx} \frac{dg_{xx}}{dr} \Dr \ln
\left[ g^{xx}g^{rr}\left(\frac{dg_{xx}}{dr}\right)^2 \right] \right)^2
\nonumber \\
&+& \left( g^{00}\frac{dg_{00}}{dr} \Dr \ln
\left[ g^{00}g^{rr}\left(\frac{dg_{00}}{dr}\right)^2 \right] \right)^2 \Bigg\} \, .
\ea

\newpage
\begin{figure}
\centerline{\epsfig{figure=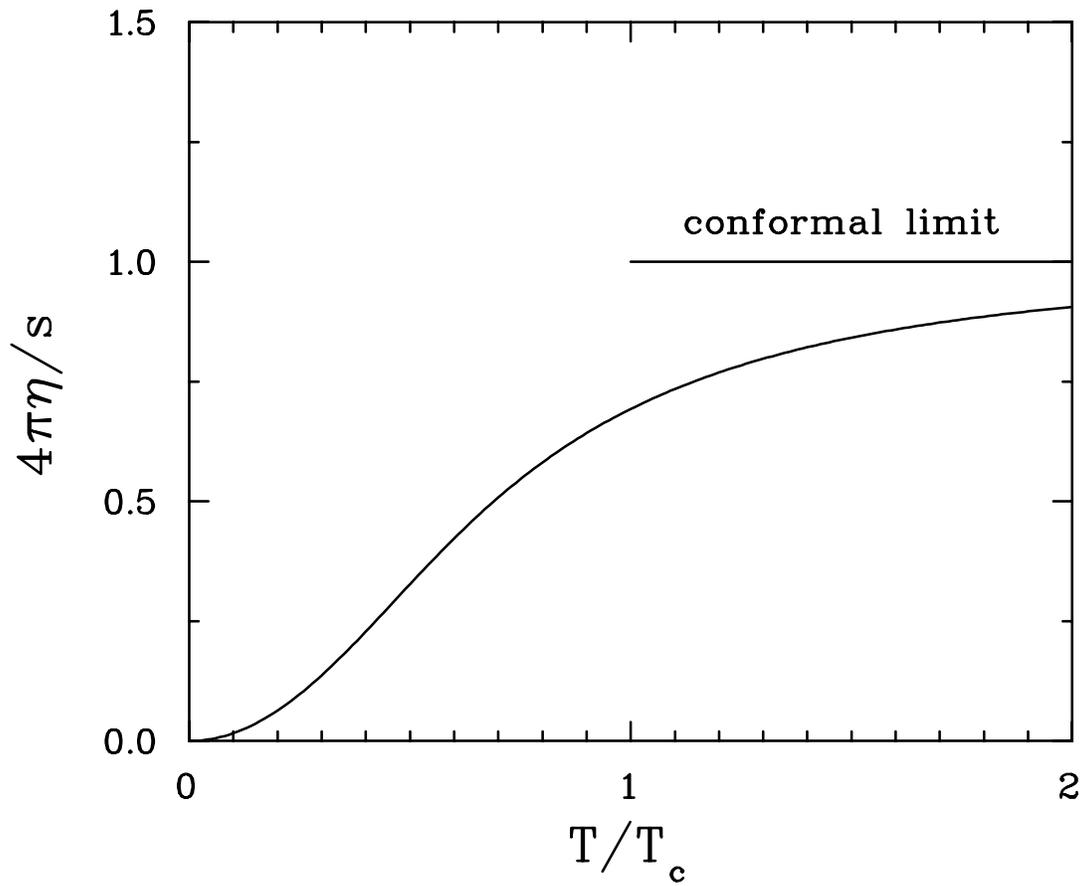,width=12.5cm,angle=90}}
\caption{The ratio of the shear viscosity to entropy density as a function of temperature in the soft wall model.}
\end{figure}

\newpage
\begin{figure}
\centerline{\epsfig{figure=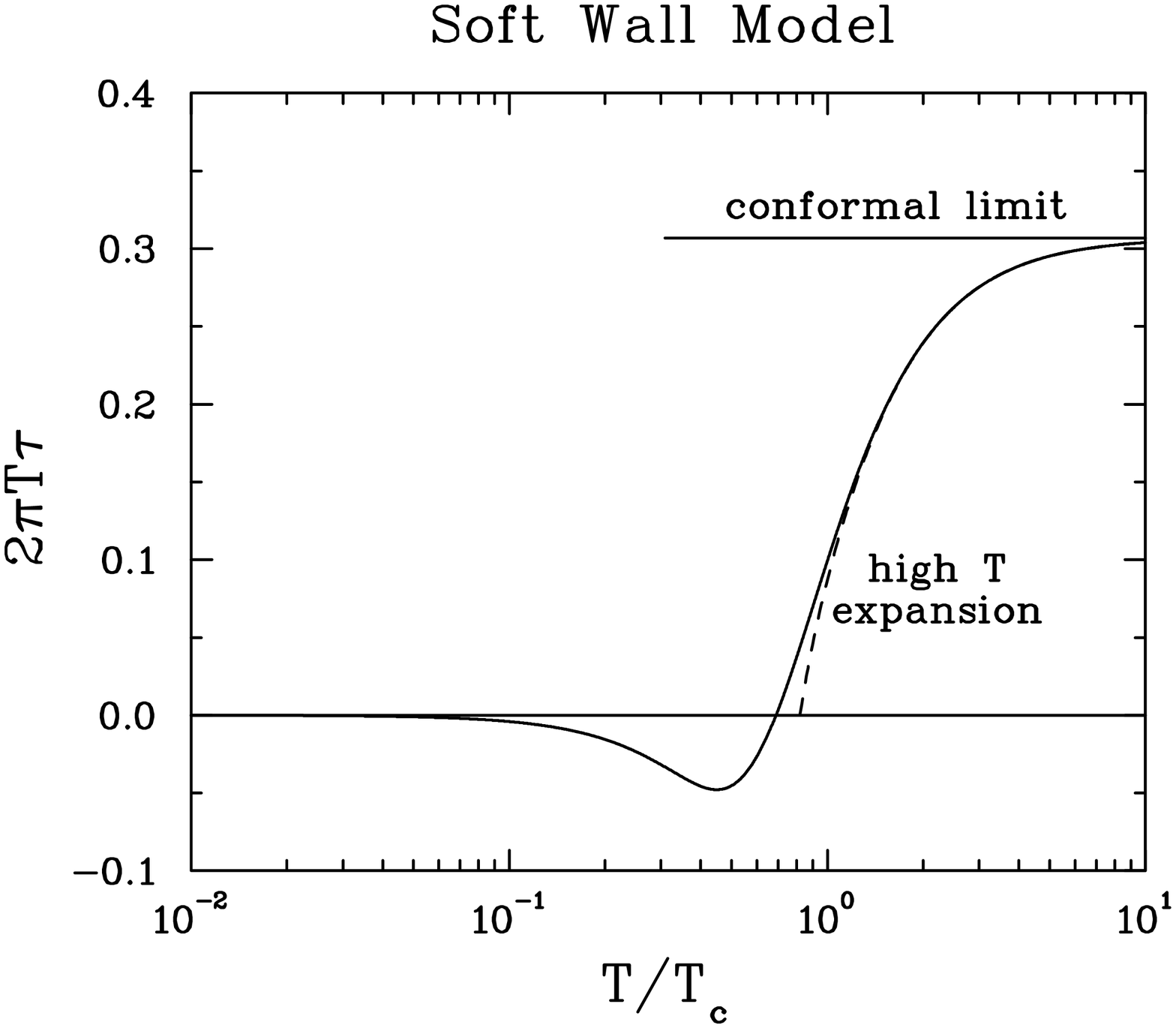,width=12.5cm,angle=90}}
\caption{The shear relaxation time as a function of temperature in the soft wall model.  The dashed curve is the high temperature expansion given in the text.}
\end{figure}


\begin{thebibliography}{99}


\bibitem{Maldacena}
J. Maldacena, Adv. Theor. Math. Phys. {\bf 2}, 231 (1998).

\bibitem{Policastro1}
G. Policastro, D. T. Son and A. O. Starinets, Phys. Rev. Lett. {\bf 87}, 081601 (2001).

\bibitem{recipe}
D. T. Son and A. O. Starinets, JHEP {\bf 09}, 042 (2002).

\bibitem{hydroI}
G. Policastro, D. T. Son and A. O. Starinets, JHEP {\bf 09}, 043 (2002).

\bibitem{hydroII}
G. Policastro, D. T. Son and A. O. Starinets, JHEP {\bf 12}, 054 (2002).

\bibitem{Herzog1}
C. P. Herzog, JHEP {\bf 12}, 26 (2002).

\bibitem{Kovtun1}
P. Kovtun and A. O. Starinets, Phys. Rev. D {\bf 72}, 086009 (2005).

\bibitem{Mas}
J. Mas and J. Tarr\'{i}o, JHEP {\bf 05}, 036 (2007).

\bibitem{stretched}
P. Kovtun, D. T. Son and A. O. Starinets, JHEP {\bf 10}, 064 (2003). 

\bibitem{Membrane}
A. O. Starinets, arXiv:0806.3797.

\bibitem{Kovtun2}
P. Kovtun and L. G. Yaffe, Phys. Rev. D {\bf 68}, 025007 (2003).

\bibitem{conformalrelax1}
M. Natsuume and T. Okamura, Phys. Rev. D {\bf 77}, 066014 (2008).

\bibitem{conformalrelax2}
R. Baier, P. Romatschke, D. T. Son, A. O. Starinets and M. Stephanov, JHEP 
{\bf 04}, 100 (2008). 

\bibitem{Heller1}
M. P. Heller and R. A. Janik, Phys. Rev. D {\bf 76}, 025027 (2007).

\bibitem{MIS}
I. M\"{u}ller, Z. Phys. {\bf 198}, 329 (1967); 
W. Israel, Ann. Phys. (N.Y.) {\bf 100}, 310 (1976);
W. Israel and J. M. Stewart, {\it ibid.} {\bf 118}, 341 (1979).

\bibitem{Weinberg}
S. Weinberg, {\it Gravitation and Cosmology: Principles and Applications of the General Theory of Relativity} (Wiley \& Sons, New York, 1972).

\bibitem{hardwall}
J. Polchinski and M. J. Strassler, Phys. Rev. Lett. {\bf 88}, 031601 (2002).

\bibitem{softwallD}
C. P. Herzog, Phys. Rev. Lett. {\bf 98}, 091601 (2007).

\bibitem{softwall}
A. Karch, E. Katz, D. T. Son and M. A. Stephanov, Phys Rev. D {\bf 74}, 015005 (2006).

\bibitem{softwallD2}
Y. Gao, W. Xu and D. Zeng, arXiv:hep-ph/0707.0817.

\bibitem{Andreev1}
O. Andreev and V. I. Zakharov, Phys. Lett. B {\bf 645}, 437 (2007).

\bibitem{Andreev2}
O. Andreev, Phys. Rev. D {\bf 73}, 107901 (2006).

\bibitem{Andreev3}
O. Andreev, Phys Rev. D {\bf 76}, 087702 (2007).

\bibitem{Liu}
A. Buchel and J. T. Liu, Phys. Rev. Lett. {\bf 93}, 090602 (2004).

\bibitem{absorption}
I. Amado, C. Hoyos-Badajoz, K. Landsteiner and S. Montero,
JHEP {\bf 09}, 057 (2007).

\end{thebibliography}
\end{document}